%% file: main.tex
\documentclass[onecolumn]{article}
\usepackage[T1]{fontenc}
\usepackage{mathptmx}

\usepackage{setspace}
\usepackage[utf8]{inputenc}
\usepackage{amsmath}
\usepackage{amssymb}
\usepackage{graphicx}
\usepackage[normalem]{ulem}
\usepackage{color,soul}
\usepackage{dblfloatfix}    

\newcommand{\ii}{\mathrm{i}}

\newcommand{\dd}{\mathrm{d}}

\usepackage[labelsep=period]{caption}

\definecolor{dark-green}{RGB}{0, 128, 0}

\makeatletter\@ifundefined{date}{}{\date{}}
\makeatother

\usepackage[
    backend=biber,
     style=science,
 ]{biblatex}
\evensidemargin\oddsidemargin \hoffset-20mm \voffset-28.4mm
\begin{document}

\title{\textbf{Anomalous topological waves in strongly amorphous scattering networks}}

\author{Zhe Zhang$^1$, Pierre Delplace$^2$, and Romain Fleury$^1$*}

\maketitle\thispagestyle{empty}

\begin{center}
    $^1$ \emph{Laboratory of Wave Engineering, School of Electrical Engineering, EPFL, Station 11, 1015 Lausanne, Switzerland}\\
    $^2$ \emph{Ens de Lyon, CNRS, Laboratoire de physique, F-69342 Lyon, France}\\
    \emph{*To whom correspondence should be addressed. Email: romain.fleury@epfl.ch}
\end{center}

\section*{Abstract}
\textbf{Topological insulators are crystalline materials that  have revolutionized our ability to control wave transport. They provide us with unidirectional channels that are immune to obstacles, defects or local disorder, and can even survive some random deformations of their crystalline structures. However, they always break down when the level of disorder or amorphism gets too large, transitioning to a topologically trivial Anderson insulating phase. Here, we demonstrate a two-dimensional amorphous topological regime that survives arbitrarily strong levels of amorphism. We implement it for electromagnetic waves in a non-reciprocal scattering network and experimentally demonstrate the existence of unidirectional edge transport in the strong amorphous limit. This edge transport is shown to be mediated by an anomalous edge state whose topological origin is evidenced by direct topological invariant measurements. Our findings extend the reach of topological physics to a new class of systems in which strong amorphism can induce, enhance and guarantee the topological edge transport instead of impeding it.}

\section*{One-Sentence Summary}
\textit{We predict and observe a novel topological photonic phase that survives arbitrarily strong levels of amorphism.}
\section*{Introduction}
\input{introduction}

\section*{Amorphous non-reciprocal networks}

\input{Part1}

\section*{Exceptional resilience of anomalous edge states to strong amorphism}
\input{Part2}
\section*{Amorphism-enhanced topological anomalous phase}
\input{Part3}

\section*{Measurement of a topological invariant in the strongly amorphous regime}

\input{Part4}

\section*{Conclusion}
\input{Conclusions}

\newpage
\section*{Figures}
\input{Figures_Maintext.tex}

\newpage
\small
\printbibliography

\section*{Acknowledgments}
We thank Lucien Jezequel and Aleksi Bossart for useful discussions. We thank Yifei Guan, Shiling Liang, and Yuqi Song for providing assistance in taking photographs, and Qiaolu Chen for help to characterize the non-reciprocal phase shifter used in the winding number measurement.

\section*{Fundings}
This work was supported by the Swiss State Secretariat for Education, Research and Innovation (SERI) under contract number MB22.00028, and the Swiss National Science foundation under the Eccellenza award 181232.

\section*{Author contributions}
Z.Z. performed the theoretical, numerical and experimental work under the supervision of R.F. and P.D. All authors contributed to the developments of the theoretical and experimental methods, to the interpretation of the results, and to writing the manuscript.

\section*{Competing interests}
The authors declare no competing interests.

\section*{Data and materials availability}
The code and data used to produce the plots within this work will be released on the repository Zenodo upon publication of this preprint.

\section*{Supplementary Materials}
Materials and Methods\\
Supplementary Text\\
Figs. S1 to S24

\end{document}

%% file: introduction.tex
\label{introduction}

Amorphous solids \cite{berthier_glass_amorphous_review_2011}, materials that do not exhibit any structural long-range order, represent the majority of the solids found on Earth. Amorphous materials such as glasses, metals, plastics and semiconductors provide unique mechanical \cite{treacy_amorphous_silicon_2012}, electrical \cite{hong_elect_amor_semi_2020} and optical \cite{eggleton_amor_photonic_2011} properties. Yet, the physics traditionally associated with periodic structures often breaks in amorphous systems. An important example is the topological classification of matter \cite{ryu_Tenfold_2010, fu_topo_cry_2011,delplace_PRS_2017}, which inherently applies to periodic insulating structures based on their intrinsic and crystalline symmetries, and has recently stimulated the exploration of fascinating phenomena, including robust energy transport in electronic \cite{ando_topo_material_review_2013,guan_Euler_class_2022} and classical wave \cite{romain_topo_acoustics_2015,afzal_AFI_2020,zhang_superior_AFI_2021,liu_topological_3D_Chern_2022} systems. Actually, the concept of topology is not fundamentally limited to systems with spatial translation symmetry, as implied by early observations of the quantum Hall effect in disordered samples \cite{huckestein_QHE_disorder_1995}, and theoretical predictions in non-commutative geometry \cite{bourne2018non,bellissard1994noncommutative}. Recently, topological edge states were shown to be robust to small structural disorder until the close of mobility gaps \cite{mansha_Amor_topo_Chern_2017,poyhonen_amorphous_topo_glass_2018,zhou_photonic_amorphous_Chern_2020,kim_Topo_amor_theo_2022,}. Recent works have theorized amorphism-induced topology in random fermionic tight-binding lattices \cite{titum_AFI_anderson_2016,agarwala_Amorphous_topo_2017,marsal_Weaire_Thorpe_model_2020,marsal_Amor_topo_theory_obstructed_2022,wang_topo_Amor_induced_high_order_2021,wang_topo_Amor_induced_2022} and quantum spin liquids \cite{cassella_amor_spin_liquid_2022}, and performed experiments with mechanical systems \cite{mitchell_amorphous_gyroscope_2018}. However, all these prior arts were limited to weak levels of amorphism, as stronger levels of amorphism unavoidably close the mobility gaps via Anderson localization, and destroy the topological phase. In addition, the non-trivial topology of amorphous phases has so far only been numerically confirmed using either local Chern markers \cite{kitaev_anyons_real_chern_2006} or Bott index \cite{agarwala_Amorphous_topo_2017}  calculations, and a direct experimental measurement of an amorphous topological invariant is still crucially missing. 

Here, we investigate, design and experimentally demonstrate a topological regime surviving arbitrarily large levels of amorphism, and prove its topological origin by direct topological invariant measurements. Starting from oriented scattering graphs, we exploit a simple non-atomic limit displaying an unidirectional boundary state to construct anomalous topological edge states that are immune to Anderson localization regardless of the level of amorphism. We implement our strategy for electromagnetic waves, by building non-reciprocal scattering networks operated in the GHz range. We experimentally show the topological edge transport in samples with maximal amorphism, by combining transmission and topological invariant measurements. Those direct measurements together with numerical simulations unveil a striking transition mechanism, in which the interplay between amorphism and topology triggers, enhances and guarantees the nucleation of robust topological wave energy transport.

%% file: Part1.tex

Scattering networks are simple and powerful models to investigate topological phenomena \cite{Ho_Chalker_percolation_1988,Ryu_spin_network_2014,chong2014network,delplace_PRS_2017,delplace_Topo_random_network_2020,upreti_floquet_2019,afzal_AFI_2020,zhang_superior_AFI_2021}. In condensed matter physics, they provide a description of the topological phase transitions in the quantum Hall \cite{Ho_Chalker_percolation_1988} and the quantum spin Hall effects \cite{Ryu_spin_network_2014}. They can represent various discrete-time quantum walks and also constitute relevant models to investigate classical wave propagation in guided circuits \cite{upreti_floquet_2019,afzal_AFI_2020,zhang_superior_AFI_2021}, e.g. in photonics or phononics. Interestingly, it was shown that planar scattering networks may not only exhibit robust unidirectional (or chiral) edge states characterized by a non-zero Chern invariant, similar to those of the quantum Hall effect, but can also display \textit{anomalous} chiral edge states that have a distinct topological origin. Such anomalous edge states were first proposed in periodically driven quantum systems \cite{kitagawa_AFI_original_2010} and discrete-time quantum walks \cite{rudner_AFI_theory_2013}, and their existence was soon after proposed in undriven periodic photonic networks \cite{chong2014network,delplace_PRS_2017,zhang_superior_AFI_2021}. 

Actually, chiral edge states can be easily understood if scattering networks are considered as oriented graphs: the wave  propagates along the links of the graph in a given direction and scatters at the nodes (Fig. \ref{fig:Concept}A, first column). If one tunes the scattering parameters of each node such that the wave is fully transmitted into only one link, one can end up with a  configuration where the wave is trapped around minimal circuits in the bulk (blue loops), but can circulate around the system through the outmost exterior links (orange loop). Such a configuration is a two-dimensional analog of the non-atomic limit of the Su–Schrieffer–Heeger (SSH) model where the atoms in the bulk of the chain pair to form uncoupled diatomic molecules, leaving an isolated single atom at each  extremity of the chain. In our case, it turns out that this surrounding circuit precisely corresponds to an ideal anomalous chiral edge state that decouples from the rest of the bulk, as opposed to more conventional  edge states of Chern phases that do not support a similar graph interpretation \cite{delplace_PRS_2017}. The existence of such unidirectional surrounding circuits is guaranteed for planar oriented graphs \cite{delplace_Topo_random_network_2020} with equal numbers of incoming and outgoing modes at each node, known as Eulerian graphs. 
Remarkably, it does not depend on the spatial arrangement of the nodes of the graph nor on the length of the links. This suggests that the anomalous chiral edge states of scattering networks are robust to arbitrarily large amorphism (Fig. \ref{fig:Concept}A, bottom left panel).


Guided by this insight, we propose a physical realization of such amorphous insulators by exploiting a complete mapping between oriented graphs and non-reciprocal networks made of circulators (Fig. \ref{fig:Concept}A, second column). The circulator networks are generated using a weighted Voronoi tessellation, picking random weights on a triangular generator set lattice (see Supplementary Materials for details). The weight standard deviation defines a dimensionless amorphous factor $\alpha$ that allows us to describe the continuous deformation from the clean-limit honeycomb lattice (top row, $\alpha=0$), to a highly amorphous regime (bottom row, $\alpha=6$). We implemented this strategy into two electromagnetic prototypes made of GHz circulators connected by microstrip lines (third column). The measured electric field maps (fourth column) confirm the existence of the anomalous edge state even when $\alpha=6$.

The level of amorphism used in this experiment is very strong, as evidenced from the evolution of network statistics with $\alpha$, shown in the left panel of Fig. \ref{fig:Concept}B. When $\alpha$ is increased from 0, the initial honeycomb network first deforms into a weakly amorphous one composed of loops with a preserved number of sides (still equal to 6), but of variable link lengths. In the range $1.75\lesssim \alpha \lesssim 5$, a transition occurs during which the percentage of 6-sided loops drops significantly as the length variance keep rising. Finally, for $\alpha \gtrsim 5$, the statistics become steady and the level of amorphism cannot be further increased, reaching the maximal value achievable on an Euclidean plane. The loss of 6-sided loops during the transition is accompanied with the creation of a stable distribution of loops with smaller and larger number of sides, as evidenced in the right panel.


%% file: Part2.tex

In order to evidence the remarkable resilience of anomalous edge states to a high level of amorphism, we now compare them with the usual edge states of the Chern phase. In the clean limit of the honeycomb lattice, trivial, Chern and anomalous phases can be engineered according to the degree of reflection and nonreciprocity of the circulators \cite{zhang_superior_AFI_2021}. We use this property to numerically investigate the fate of wave propagation along anomalous-trivial and Chern-trivial interfaces, comparing the clean case to the strong amorphous limit (Fig. \ref{fig:interface}A).
While in the clean limit ($\alpha=0$, top row), the two configurations both display the expected  chiral mode along the interface, in the strong amorphous regime ($\alpha=8$, bottom row) the anomalous edge state is the only one to survive. The Chern-trivial interface configuration no longer benefits from a topological protection, as the input wave has localized. In other words, both topologically trivial and Chern insulators transit toward a topologically trivial Anderson insulator in the strongly amorphous regime, in sharp contrast with the anomalous phase, suggesting the existence of a topological amorphous phase immune to Anderson localization. We confirm experimentally the topological distinction between the two amorphous phases, respectively obtained  by building an amorphous network in the regime $\alpha=6$, made of two domains that respectively correspond to Chern and anomalous phases in the clean limit (Fig. \ref{fig:interface}B). 
We observe the unidirectional propagation of an interface state, from the top to the bottom (Fig. \ref{fig:interface}C for an excitation from the top, see Supplementary Materials for the excitation from the bottom), in agreement with a difference of topology between the two strongly amorphous networks.

%% file: Part3.tex

To shed light on how this amorphous anomalous  phase emerges from the clean limit, we compute the evolution of the transmission coefficient between two ports of a finite anomalous network when increasing  the amorphous factor $\alpha$ (Fig. \ref{fig:transmission}A). The ports are located either both on the edge (top) or both inside the bulk (bottom). 

In the clean limit ($\alpha=0$), the bulk band structure is well defined in terms of the link phase delay $\varphi$ between the nodes whose possible values form bands. For a given value of $\varphi$, the network  is either in a topological gap with a chiral edge state or in a bulk band (blue and light orange regions near the vertical axes in Fig. \ref{fig:transmission}A, respectively). In the weakly amorphous range, the transmission remains consistent with that of the clean limit: edge transmission is always unity in the anomalous topological gaps, and fluctuates within bands, whereas bulk transmission vanishes in the gaps but remains finite in the bulk bands. This reminiscence of the clean limit band structure completely disappears in the fully amorphous regime, which is characterized by unitary edge transmission and zero bulk transmission, regardless of $\varphi$. 
This shows that strong amorphism drives a global enhancement of the edge transmission and of the bulk insulation when compared to the clean  limit. This is markedly different from the behavior of trivial and Chern insulators, which both undergo a trivial Anderson localization transition, both with nearly zero edge and bulk transmissions (see Supplementary Materials).

The transition of the bulk bands into topological amorphous insulators is by no mean trivial. Fixing $\varphi=1.1$, we now track the evolution of the excited field by plotting its distribution at selected values of $\alpha$ (Fig. \ref{fig:transmission}B). In the clean limit (top row), the field is delocalized regardless of the position of the ports, consistently with the existence of a bulk band. This clean limit picture remains valid when $\alpha$ is increased a little, within the weakly amorphous regime. For $\alpha=2$, we reach a transition where the bulk transmission has been reduced and the field map shows that the states excited by the bulk and edge probes are much less delocalized. Keeping increasing $\alpha$, the field excited in the bulk localizes around the input port, whereas the edge scenario exhibits a robust chiral edge channel, with enhanced edge localization under strong amorphism.

%% file: Part4.tex
We now provide a direct experimental evidence  of the topological character of the anomalous amorphous insulating phase. We start from a finite strongly amorphous network and endow it with a twisted boundary condition on its lateral sides, effectively wrapping it up into a cylinder (Fig. \ref{fig:invariant}A). The twisted boundary condition imposes a direction-dependent phase delay ($\Phi$ from left to right, and -$\Phi$ in reverse), which plays, in this photonic system, the same role as the synthetic magnetic flux $\Phi$ threading the cylindrical electronic sample considered in Laughlin’s thought experiment \cite{laughlin_quantized_original_1981,hafezi_measuring_2014,hu_measurement_Edge_invariant_2015,fabre_laughlins_Exp_2022}.
The system is then probed through one external port at the bottom edge. The topological invariant \textit{W} is defined as the winding number of the probe's reflection coefficient $R$ when $\Phi$ is adiabatically varied from $0$ to $2 \pi$ \cite{braunlich2010equivalence,fulga2012scattering,meidan_topo_scattering_invariant_2011}, that is $W=\frac{1}{2 \pi \ii}\int_0^{2\pi} \, \dd \Phi \, R^* \frac{\partial R}{\partial \Phi}$.
To implement the twisted boundary conditions experimentally in our photonic device, we have conceived a non-reciprocal phase shifter (Fig. \ref{fig:invariant}B), in which the direction-dependent phase delays are set by external voltages $\textit{V}_1$ and $\textit{V}_2$ (see details in Supplementary Materials).

Figure \ref{fig:invariant}C reports the measurement of the reflection coefficient's winding for the anomalous phase in the clean limit, and compares it with the one obtained in the strongly amorphous regime ($\alpha=6$). As a reference, we also performed a measurement starting from a Chern insulator. For clarity, we plot as well the field maps measured when disconnecting the twisted boundary condition. Whereas both anomalous and Chern phases do wind in the clean limit, only the anomalous phase shows a nonzero winding when strong amorphism is present. Such non-zero winding is always accompanied with edge transport. In Fig. \ref{fig:invariant}D, we repeat the experiment but starting from non-insulating samples. In the clean limit, \textit{W} is zero in both cases as expected for bulk bands. Turning on strong amorphism, the Chern sample remains topologically trivial, whereas we now measure a non-zero winding for the anomalous one. Our measurements prove the topological origin of the edge states, and the emergence of a topological phase under very strong amorphism. 

While in practice we can only explore a few particular realizations, numerical simulations can reveal the fascinating topological transition driven by strong amorphism. Fig. \ref{fig:invariant}E shows the numerically calculated winding numbers of randomly generated networks for all levels of amorphism and values of $\varphi$. For anomalous networks, the phase diagram suggests that the effect of amorphism is first to localize the bulk states, creating a topological mobility gap. This is confirmed by participation ratio calculations (see Supplementary Materials). Besides, we also find that the anomalous spectrum in the cylinder geometry displays flat bulk bands with gapless chiral edge states in between. In contrast, when starting from a Chern phase, the topological gaps close as both bulk and edge states localize, creating a trivial amorphous insulator. Performing a statistical study at $\varphi=1.1$ with 200 realizations, Fig. \ref{fig:invariant}F shows that topological transitions have to be understood in a statistical sense: the proportion of realizations with nontrivial winding number changes smoothly when increasing $\alpha$ through the transition stage.

%% file: Conclusions.tex
We have reported on the observation of an anomalous non-crystalline topological phase that survives arbitrarily large levels of amorphism, escaping Anderson localization. We explained intuitively the robustness of such a phase using graph theory arguments, and implemented it in a microwave experiment allowing for a direct observation of edge states. We proposed a practical method to perform direct topological invariant measurements in finite aperiodic samples, and shed light on the complex localization mechanisms involved in the nucleation of anomalous amorphous topological insulators. We envision a new class of topological systems in which strong amorphism is no longer a hindrance, but can be used as a new degree of freedom to induce, control and strengthen the robustness of unidirectional wave energy transport.

%% file: Figures_Maintext.tex
\begin{figure*}[!h]
  \centering
  \includegraphics[width =0.95 \columnwidth]{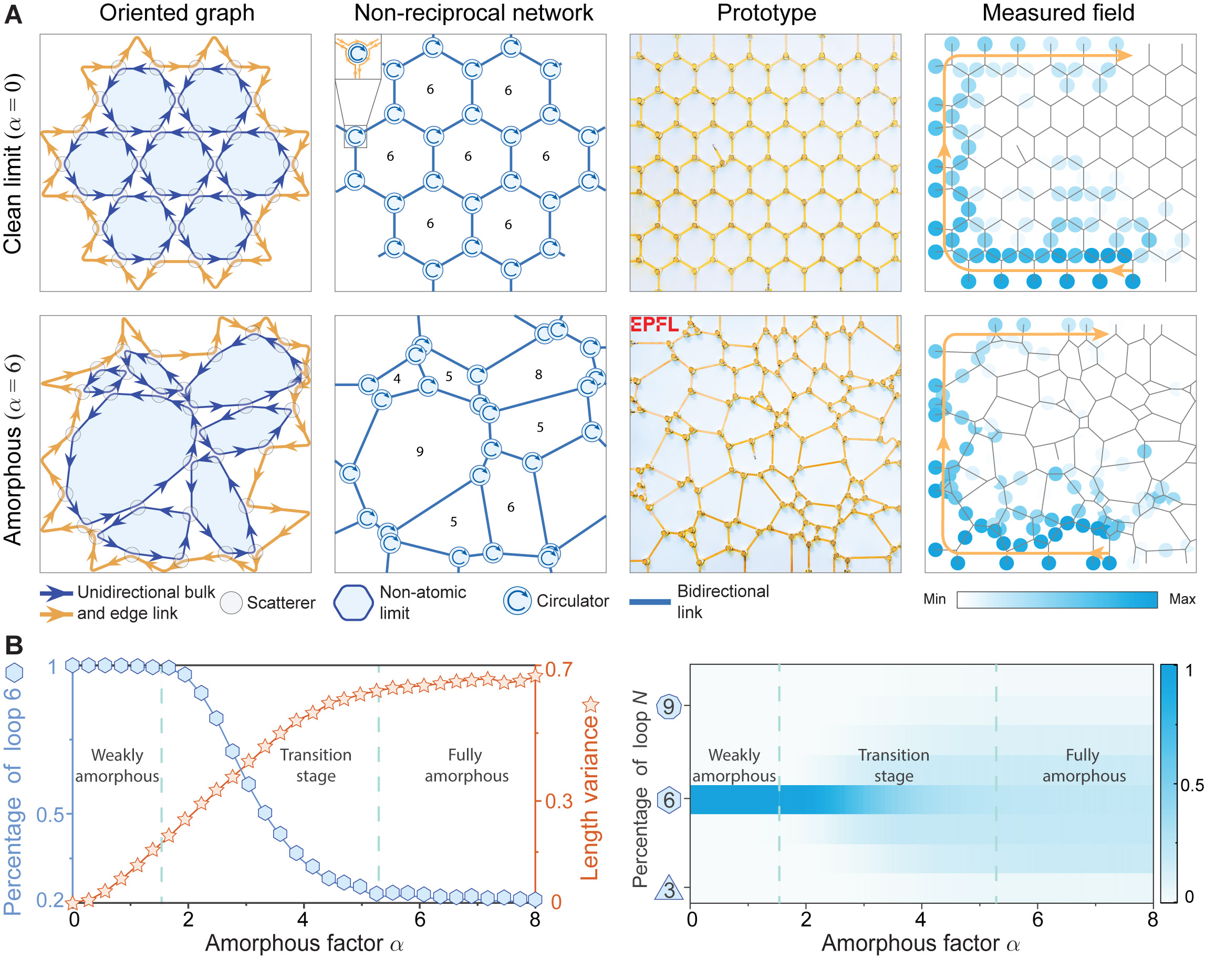}
  \caption{
  \textbf{Anomalous edge states can survive any level of amorphism.}
  (\textbf{A}) Anomalous topological edge states occur in any scattering signal graphs, when a limit can be found in which bulk signals travel in closed loops (in blue), leaving a large signal loop on the edge (in orange). This picture is valid not only for periodic systems (top row, honeycomb case), but should also be true for any level of amorphism (bottom row). We validate this idea by mapping the oriented graphs (first column) to practical scattering networks made of three-port circulators linked with reciprocal connections (second column). We built prototypes (third column) operating around 5.7 GHz and experimentally observed the resilience of anomalous edge states to strong levels of amorphism (fourth column).
  (\textbf{B}) To characterize the effect of amorphism, we introduce an amorphous factor $\alpha$ that controls the continuous transition from the clean limit to a strongly amorphous phase. We track both the link length variance and the number of sides of each loop in the network (numbers in panel a, second column). As $\alpha$ increases from 0 to 1.75, the links start to deform but the percentage of loops with $N = 6$ sides stays at $100\%$ as in the clean honeycomb limit (left panel). Beyond this weakly amorphous regime, a transition occurs during which the percentage of loop 6 drops significantly as the one of loops with $N \neq 6$ increases. For $\alpha$ above 5, we enter a fully amorphous phase, characterized by a stabilized distribution of loops of various sizes. The networks statistics are computed from 1000 random realizations of networks made of 1000 nodes.
  }
  \label{fig:Concept}
\end{figure*}
\newpage

\newpage
\begin{figure*}[!h]
  \centering
  \includegraphics[width =0.95 \columnwidth]{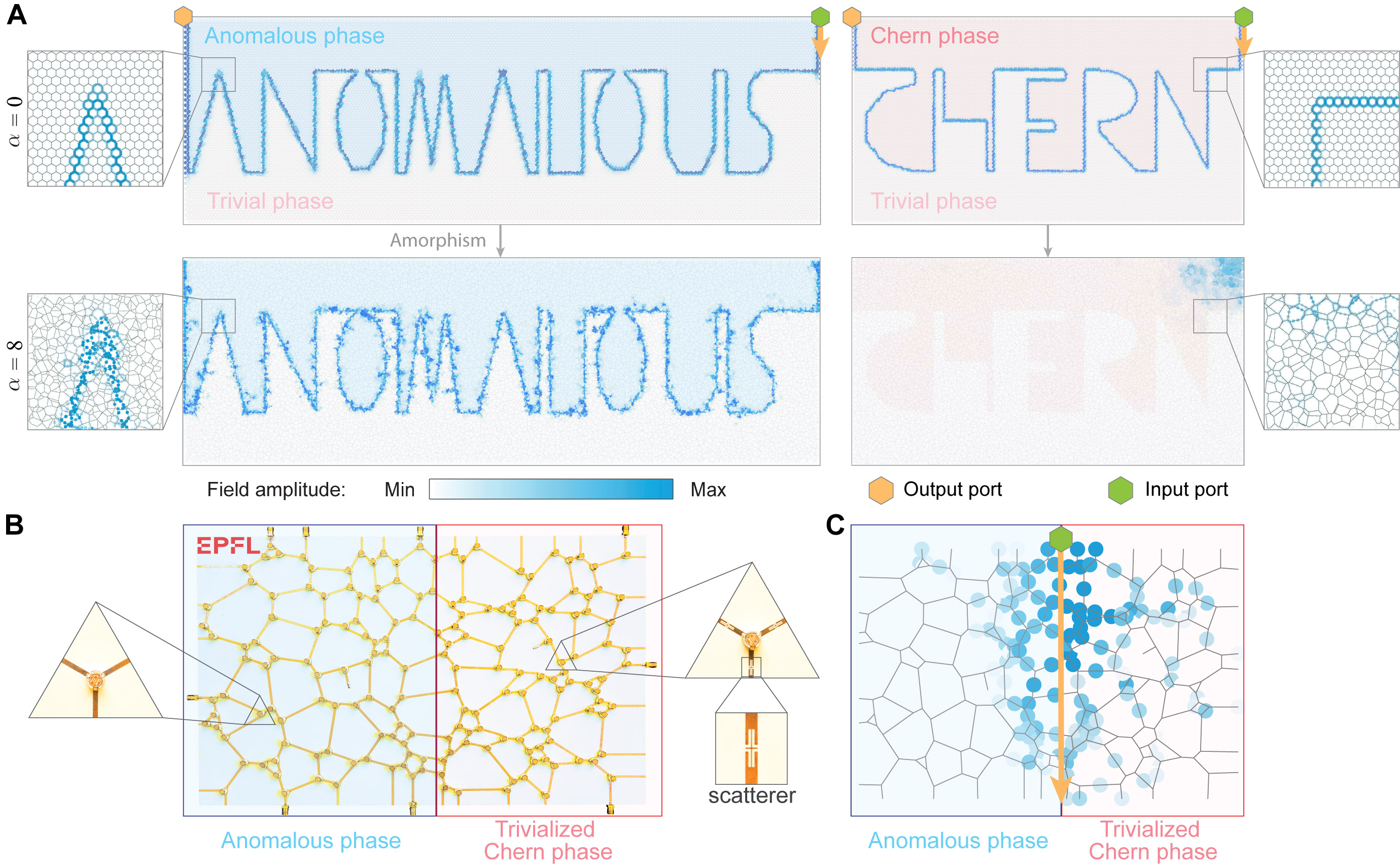}
  \caption{
  \textbf{ Exceptional resilience of anomalous edge states to strong amorphism.} We compare the resilience of the anomalous edge mode to amorphism with the one of a standard Chern edge mode, when propagating along a domain wall with a trivial insulator. (\textbf{A}) In the clean limit ($\alpha = 0$, first row), both anomalous and Chern phases provide a robust   channel with unitary transmission. Then, we impart strong amorphism ($\alpha = 8$, second row). Only the anomalous interface state survives. Conversely, the Chern case undergoes Anderson localization.
  (\textbf{B}) Experimental demonstration of the topological distinction between anomalous (left) and trivialized Chern (right) phases in the strong amorphous limit ($\alpha = 6$). The trivialized Chern phase is obtained by adding amorphism to a Chern crystal, which differs from an anomalous phase only by the presence of extra scatterers between the circulators. 
  (\textbf{C}) Measured field map when exciting the interface from the top, confirming the existence of a topological state at the interface.}
  \label{fig:interface}
\end{figure*}
\newpage

\newpage
\begin{figure*}[!h]
  \centering
  \includegraphics[width =0.75 \columnwidth]{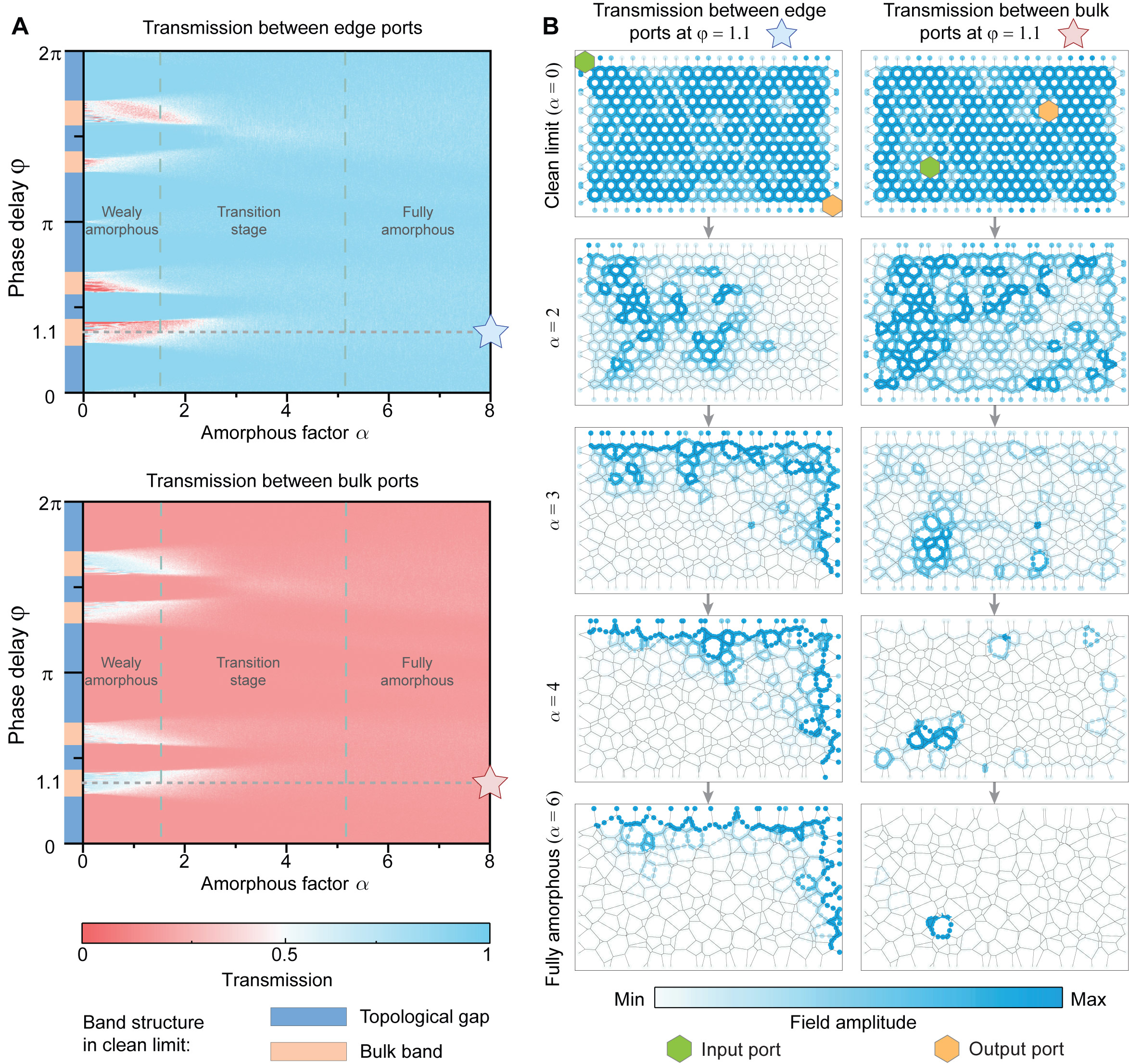}
  \caption{
  \textbf{Amorphism-enhanced edge transmission and bulk insulation of the anomalous phase.}
  We consider the evolution of the anomalous edge and bulk transmissions when increasing the level of amorphism, for any value of the phase delay $\varphi \in [0, 2\pi]$ of the reciprocal links, defined in the clean limit. Each point corresponds to an average over 200 realizations of randomly generated scattering networks, with some examples shown in panel (B). (\textbf{A}) In the weakly amorphous regime, the edge (top) and bulk (bottom) transmissions are consistent with the clean limit band structures, with large edge transmission only in the topological gaps, and non-zero bulk transmission only in the bulk bands. After the transition stage, the edge transmission is enhanced to 1 and the bulk transmission is pinned to zero regardless of the value of $\varphi$. This confirms the nucleation of a single amorphism-enhanced anomalous topological phase, which now spans the full $2\pi$ range.
(\textbf{B}) Examples of amorphous networks and simulated fields of edge and bulk transmissions at $\varphi = 1.1$ (dashed line marked with stars in panel (A)), which corresponds to a bulk band in the clean limit. The panels demonstrate how the bulk modes localize as the amorphous factor is increased from the clean limit (first row) to the fully amorphous case (last row), creating an amorphous topological phase with large edge transmission.}
  \label{fig:transmission}
\end{figure*}
\newpage

\newpage
\begin{figure*}[!h]
  \centering
  \includegraphics[width =0.95 \columnwidth]{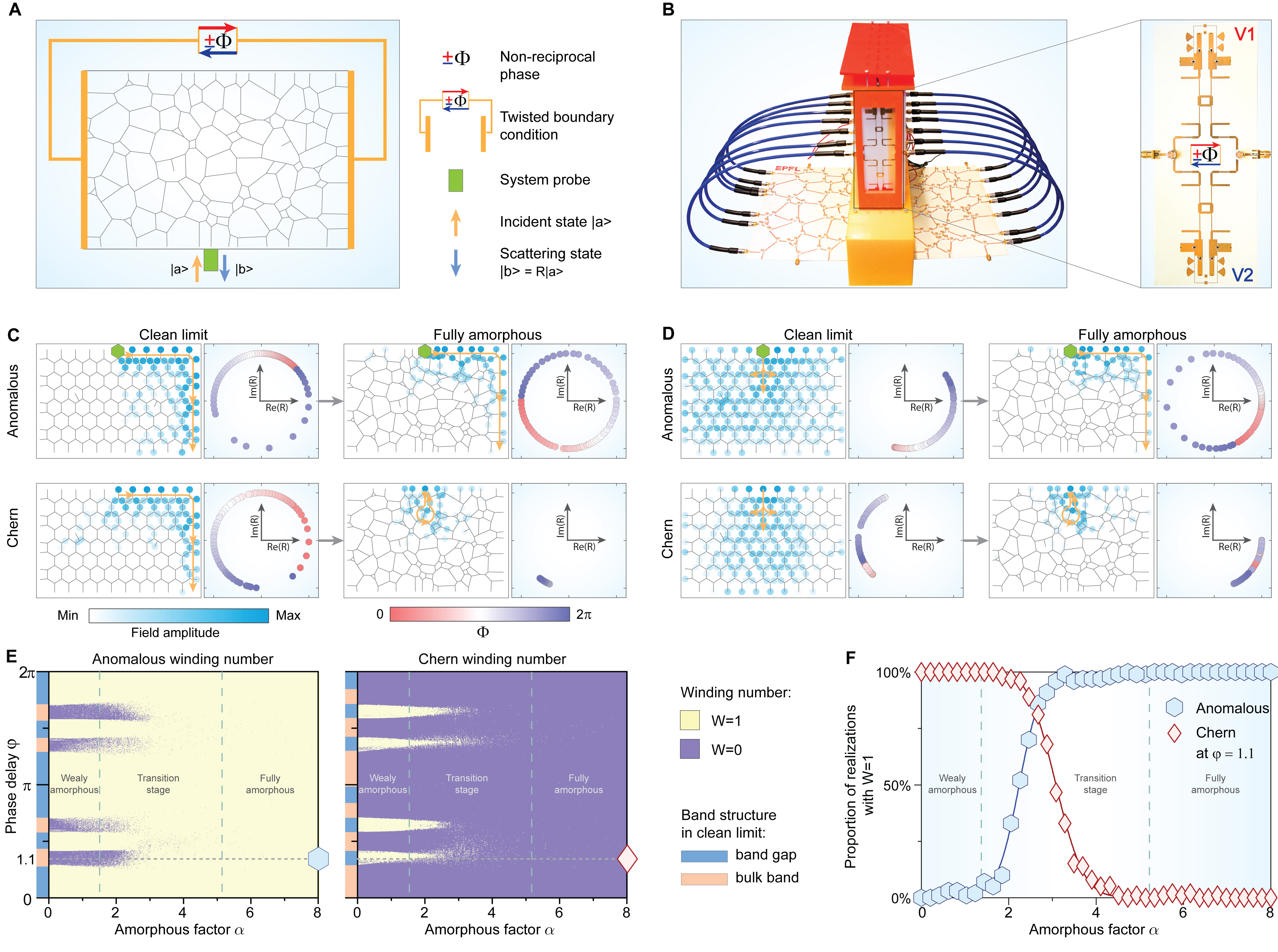}
  \caption{
  \textbf{Direct measurement of topological indices in the strongly amorphous regime.}
(\textbf{A}) Scheme for measuring the topological index of amorphous scattering networks. We impart a twisted boundary condition to the scattering network, wrapping up the sample into a cylinder. The boundary condition links the left and right boundaries with a non-reciprocal phase $\Phi$. The topological index \textit{W} is the winding of the reflection coefficient $R$ measured at the external probe, when $\Phi$ is varied over all angles.
(\textbf{B}) Picture of the experimental setup, with the microwave non-reciprocal phase shifter shown in the inset. The value of $\Phi$ is controlled by external d.c. voltages $V_1$ and $V_2$.
(\textbf{C} and \textbf{D}) Measured winding numbers and corresponding field maps, when starting from different situations in the clean limit. In panel (C), we start with anomalous and Chern networks in a topological gap, whereas in panel (D), we start inside bulk bands. The measurements show that regardless of the starting point in the clean limit, the anomalous network is always topological under strong amorphism. Conversely, the Chern network always becomes a trivial insulator.
 (\textbf{E}) Amorphism-induced topological phase transitions. For each amorphous level $\alpha$ and phase delay $\varphi$, we calculate the winding number \textit{W} of a randomly generated network. In the weakly amorphous regime, the bulk bands of the anomalous network persist with trivial windings, while the Chern networks exhibit a robust non-trivial topology only near the Chern gaps already present in the clean limit. Under moderate levels of amorphism, opposite transitions occur for the two phases. The occurrence of Chern networks with non-zero winding decreases to zero, consistent with a trivial Anderson insulator. Very differently, anomalous networks undergo a topologically nontrivial Anderson transition, as the entire spectrum becomes topological.
 (\textbf{F}) Statistical study of the proportion of realizations with non-trivial winding versus $\alpha$, performed on 200 random realizations of amorphous networks with phase delay $\varphi = 1.1$. This confirms the opposite topological transitions occurring at moderate amorphism levels for both the Chern and anomalous cases, consistent with our measurements in panels (C) (Chern) and (D) (anomalous).
  }
  \label{fig:invariant}
\end{figure*}